\documentclass{article}
\usepackage{graphicx}	
\usepackage{bm}
\usepackage{amsmath}    
\usepackage{amssymb}
\usepackage{mathtools}
\usepackage{xcolor}
\usepackage{epstopdf}
\usepackage{upgreek}
\usepackage{subfig}
\usepackage{natbib}
\usepackage{authblk}

\def \E{{\rm E}}
\newcommand{\mT}{\ensuremath{\mathsf{T}}}

\begin{document}

\title{A hierarchical model for estimating exposure-response curves from multiple studies}

\author[1]{{Joshua P.} {Keller}}
\author[2]{{Joanne} {Katz}}
\author[3]{{Amod K.} {Pokhrel}}
\author[3]{{Michael N.} {Bates}}
\author[4]{{James} {Tielsch}}
\author[2]{{Scott L.} {Zeger}}
\affil[1]{Colorado State University}
\affil[2]{Johns Hopkins University}
\affil[3]{University of California, Berkeley}
\affil[4]{George Washington University}

\renewcommand{\thefootnote}{}
\footnotetext{\noindent Corresponding author: Joshua P. Keller, Department of Statistics, Colorado State University, Fort Collins, CO USA.
Email: {joshua.keller@colostate.edu}.}

\date{}

\maketitle

\begin{abstract}
{Cookstove replacement trials have found mixed results on their impact on respiratory health. The limited range of concentrations and small sample sizes of individual studies are important factors that may be limiting their statistical power. 
We present a hierarchical approach to modeling exposure concentrations and 
pooling data from multiple studies in order to estimate a common exposure-response curve. 
The exposure concentration model accommodates temporally sparse, clustered longitudinal observations. The exposure-response curve model  provides a flexible, semi-parametric estimate of the exposure-response relationship while accommodating heterogeneous clustered data. We apply this model to data from three studies of cookstoves and respiratory infections in children in Nepal, which represent three study types: crossover trial, parallel trial, and case-control study. We find evidence of increased  odds of disease for particulate matter concentrations between 50 and 200 $\upmu$g/m$^3$ and a flattening of the exposure-response curve for higher exposure concentrations. The model we present can incorporate additional studies and be applied to other settings.}
\end{abstract}

\section{Introduction}
Indoor air pollution from cookstoves constitutes a major source of morbidity and mortality throughout the world \citep{Gordon2014, Mortimer2012}. Particulate matter emitted from stoves can impact respiratory and cardiovascular health of household residents, who are typically exposed to these pollutants on a daily basis. Children are a particularly vulnerable population that can be adversely impacted by indoor air pollution. Acute lower respiratory infections (ALRIs) present a significant morbidity burden that has been associated with indoor air pollution from cookstoves.

Replacing biomass burning stoves with other, cleaner-burning stoves has been proposed as a means to mitigate this health burden \citep{Grieshop2011}. Multiple studies conducted in various countries around the world have attempted to assess the health impact of replacing existing stoves with propane-fueled stoves or other cleaner technology \citep{Smith2011,Mortimer2017, Tielsch2014}. However, many of these studies have failed to identify evidence of a significant impact on respiratory disease risk attributable to stove replacement, and better evidence about the effectiveness of cookstove interventions is needed \citep{Clark2013,Rosenthal2015}. 
While cleaner-burning stoves may reduce indoor PM concentrations by a large absolute or relative amount \citep{Steenland2018}, the concentrations can still be quite high compared to the World Health Organization (WHO) air quality standard of 10 $\upmu$g/m$^3$. Furthermore, the use of multiple stove technologies (``stove stacking'') is widespread \citep{Ruiz-Mercado2015}, so analyses based upon actual concentrations rather than assigned stoves types are needed. 
Pooling data from multiple trials is one promising approach to address these concerns by using information from multiple ranges of exposure in different settings.

As part of the Global Burden of Disease, \cite{Burnett2014} developed an `integrated` exposure response (IER) curve for fine particulate matter (PM$_{2.5}$) exposure and risk of ALRI that included data from ambient air pollution studies, household air pollution studies, and second-hand smoking studies. However, major limitations of the \cite{Burnett2014} IER curve were that only a single study of household air pollution and ALRI was available and that it used a fully parametric form.  
A recent analysis of a data from a case-control study in Bhaktapur, Nepal by \cite{Bates2018} used flexible splines to estimate a similar curve with fewer assumptions, but still included data from only one study.

In this manuscript, we present a model for combining data from multiple studies in a consistent, yet flexible manner to estimate an exposure-response curve. We do this by developing hierarchical models for  long-term exposure concentrations and  for disease risk. While motivated by cookstove studies of air pollution and ALRI, the model we present can be applied to many other settings. In Section~\ref{sec:motivating_studies} we describe  three studies from Nepal that 
motivated model development.
We present our hierarchical model for exposure concentrations in Section~\ref{sec:expmodel} and the  outcome model in Section~\ref{sec:outmodel}. 
We demonstrate the model using simulations in Section~\ref{sec:simulation} and in Section~\ref{sec:application_to_nepal} we apply these models to the Nepal studies.

\section{Motivating Studies}
\label{sec:motivating_studies}

\subsection{Example study I: Bhaktapur}
\begin{figure}
\begin{center}
\includegraphics[width=0.9\textwidth]{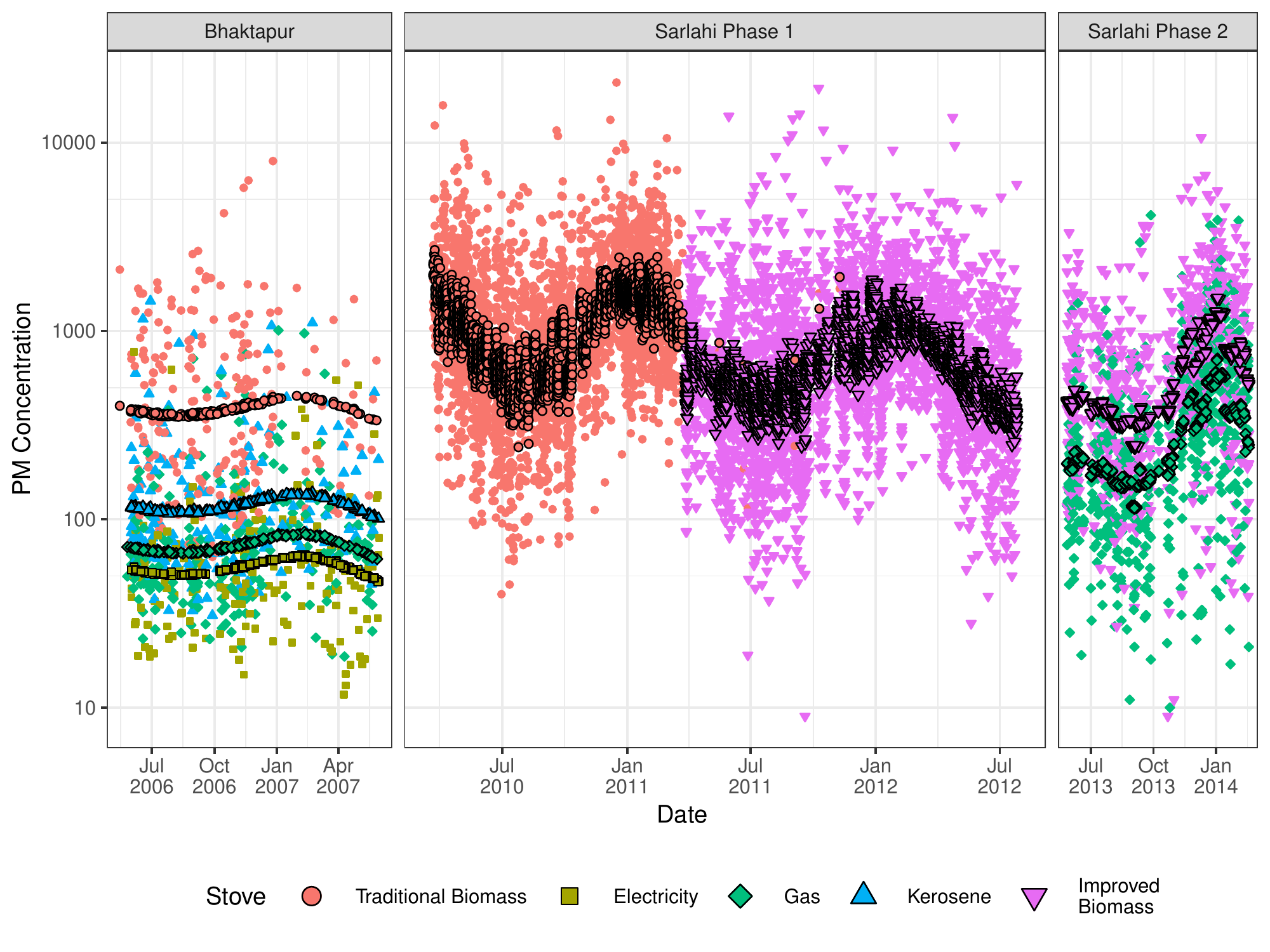}
\caption{PM$_{2.5}$ measurements (in $\upmu$g/m$^3$) from the three motivating studies (points without outlines) and fitted values from the model (outlined points).}
\label{fig:expdata}
\end{center}
\end{figure}

\begin{figure}[th]
\begin{center}
\includegraphics[width=0.9\textwidth]{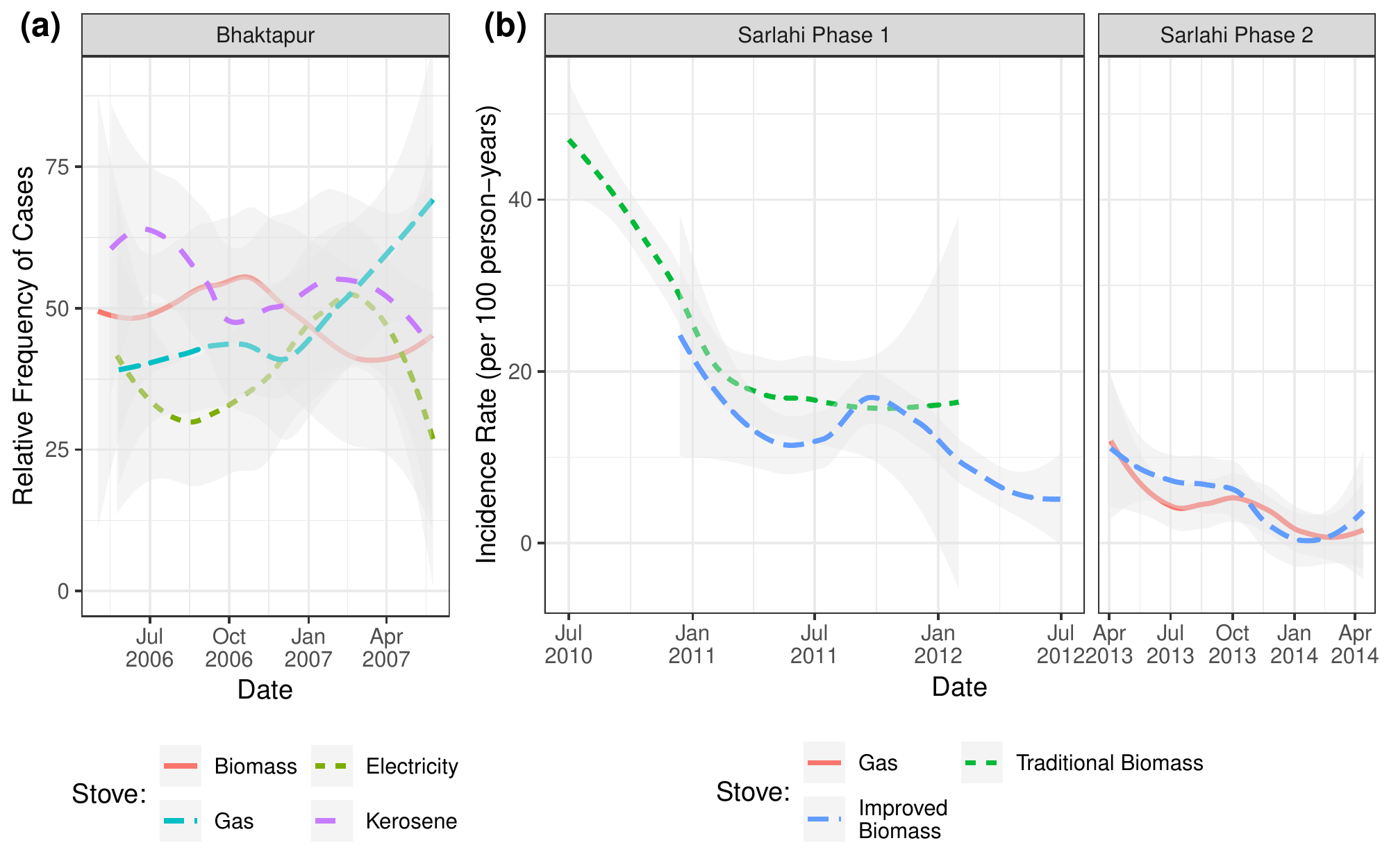}
\caption{ALRI rates across time for the (a) Bhaktapur study and (b) the two phases of the Sarlahi study. The case-control design of the Bhaktapur means that only relative rates of cases are plotted for that study, while incidence rates are shown for the Sarlahi study.} 
\label{fig:morb_comb}
\end{center}
\end{figure}

The first example study we consider is a case-control study conducted in Bhaktapur, Nepal \citep{Bates2013,Bates2018}. The study included active surveillance for respiratory illness in an open cohort of approximately 4500 children under 3 from May 2006 through June 2007. There were a total of 393  cases  who were assessed to have  ALRI according to WHO criteria \citep{Pokhrel2015}.
Control subjects were selected from the case-free subset of the cohort immediately after a case occurred. 
PM$_{2.5}$  concentrations were measured  in each child's home within a week of identification as a case or selection as a control. Stove type was identified by a questionnaire;  there were 218 households using biomass stoves, 187 using kerosene, 238 using gas, and 181 using electricity. The PM$_{2.5}$ measurements are plotted in Figure~\ref{fig:expdata} by stove type.
Figure~\ref{fig:morb_comb} shows the relative rates of cases in each stove group over time.

\subsection{Example studies II \& III: Sarlahi}
The next example studies we consider are from a cookstove intervention trial conducted in the Sarlahi district of Nepal \citep{Tielsch2014}.
This trial comprised two phases, which we  treat here as separate studies for the purpose of illustration. The first phase was a step-wedge design with  3,276 households that compared the impact of replacing traditional biomass stoves with improved biomass stoves. The second phase used a parallel design, covering a shorter enrollment period and smaller cohort of 1,312 households, to compare the improved biomass stoves against liquid propane gas (LPG) stoves.
Measurements of PM$_{2.5}$ were made in each household twice during the Phase 1 study (one measurement before and one measurement after stove installation) and once per household in Phase 2 (see Figure~\ref{fig:expdata}). In Phase 1,  there were 3,075 measurements from household using traditional biomass stoves and 2,836 PM measurements from households after switching to the improved biomass stoves.  
The geometric mean measured concentration was 1000 $\upmu$g/m$^3$ from homes using traditional stoves  and 626 $\upmu$g/m$^3$ for homes after switching to improved biomass stove. 
In Phase 2, 
there were PM$_{2.5}$ measurements for 654 and 658 households using the improved biomass stove and LPG, respectively (Figure~\ref{fig:expdata}).  The geometric mean measured value among biomass-stove households was 555 $\upmu$g/m$^3$ and in LPG households it was 265 $\upmu$g/m$^3$.

In both phases, weekly follow-up was conducted with each family to ascertain incidence of ALRI, using WHO criteria \citep{Tielsch2014}. 
In Phase 1 of the Sarlahi study, there were 598 and 217 ALRI cases among the traditional and improved biomass stove groups, respectively, yielding unadjusted rates of 29.0 and 10.5 cases per 100 person-years. As can be seen in Figure~\ref{fig:morb_comb}, these differences are impacted by strong temporal trends in ALRI incidence.  Phase 2 of the Sarlahi study had a limited number of ALRI cases, 27 and 22 in the biomass and LPG stove groups, respectively.

\section{Exposure Concentration Model}
\label{sec:expmodel}
We first present a hierarchical Bayesian model for the exposure concentrations. While we use the language of  indoor air pollution and cookstove interventions from the motivating studies, this model could be applied to a broad range of settings with continuous exposure. The key features of the study design that this model is designed to address are (a) longitudinal data with a smooth temporal trend, (b) a limited number of observations per study unit, and (c) clustering of observations.
We use a common hierarchical exposure model structure for all studies, but fit a model to the data from each study separately. This prevents issues of exposure model feedback between studies, which can differ greatly in instruments, exposure ranges, measurement procedures, behavioral patterns, and other factors. 

Let $w_{gkit}$  be the log-transformed, measured pollutant concentration in household $i$ of cluster $k$ in stove group $g$ on day $t$.  
We model the pollutant observations $w_{gkit}$ as samples from a normal distribution $N(\mu_{gkit}^E, \sigma_{w}^2)$. The mean $\mu_{gkit}^E$ is modeled as a long-term household-level concentration ($\eta_{gki}$) and a time-varying component:
$\mu^E_{gkit} = \eta_{gki} + {\bm f}(\tau_t,df)^\mT\boldsymbol\theta.$

The $\eta_{gki}$ term represents the mean concentration across the study duration for household $i$  of cluster $k$ when in stove group $g$. 
This is a combination of a group-level mean $\eta_g$, a cluster random effect $\alpha_{0k}$, and a household-level random effect $\alpha_{1i}$. 
The household random effect $\alpha_{1i} \sim N(0, \sigma_H^2)$  accounts for correlation in repeated observations within the same household.  Similarly, the cluster random effect is distributed $\alpha_{0k} \sim N(0, \sigma_K^2)$ and accounts for neighborhood-level effects, which can reflect practices and long-term local climate.
Together, this forms the following hierarchical model for the exposure data:
\begin{align}
\label{eq:full_exp_model}
 w_{gkit} &\sim N\left(\eta_{gki} + {\bm f}(\tau_t, df)^\mT\boldsymbol\theta, \sigma^2_{W}\right) \\
\notag \eta_{gki} &=  \eta_{g} + \alpha_{0k} + \alpha_{1i} \\
\notag  \alpha_{1i} &\sim N\left(0, \sigma_{H}^2\right)\\
\notag  \alpha_{0k} &\sim N\left(0, \sigma_{K}^2\right)\\
\notag \eta_{g} &\sim N(\eta_{0},\sigma_{G}^2)\\
\notag \boldsymbol\theta &\sim N(\boldsymbol\theta_{0}, \sigma_\theta^2 I).
\end{align}

The structure of this hierarchical model shrinks the observations made in each household to their cluster-level and group-level means. Given the large amount of variation present in the single day measurements, this shrinkage can improve the stability of estimates of long-term average concentrations, in place of using just the observed values.

In the exposure model \eqref{eq:full_exp_model}, $\tau_t$ is the model time step, typically a week or month, that contains day $t$. The function ${\bm f}(\tau_t, df)$  maps the model time $\tau_t$ into a $df$-length  vector of spline coefficients. 
The coarse model time step $\tau$ allows for a broad temporal trend in the household mean concentrations that can accommodate seasonality and be fit to  data from settings in which the exact measurement date cannot be released, e.g. for privacy reasons.  
In the analyses presented here, we use a centered B-spline basis for natural cubic splines (with intercept removed), which allows for flexibility across the full range of data. For a given choice of degrees of freedom ($df$), we select knot locations using quantiles across the range of observed dates.
  The spline coefficients $\boldsymbol\theta$ are modeled as $N(\boldsymbol\theta_0, \sigma_\theta^2I)$.
For the variance terms $\sigma_j^2$, $j \in \{W, H, K, G, \theta\}$, we assign  half-normal prior distributions $\sigma_j \sim N_+(m_j,v_j)$ that are designed to be weakly informative.

To quantify the amount of shrinkage in the exposure model, we employ the \emph{pooling factors} introduced by \cite{Gelman2006}. 
Details on the calculation of pooling factors for  model \eqref{eq:full_exp_model} are provided in Section 1 of the Supplementary Materials.

\section{Outcome Model}
\label{sec:outmodel}
We now describe the outcome model that combines data from all studies to estimate an exposure-response curve. As with the exposure model, we use the language of ALRIs and air pollution because of the motivating studies, but this statistical model could be applied to a wide variety of exposure-response problems.

Let $Y_{it}$ and $T_{it}$ denote the number of a cases and total time at risk for individual $i$ during period $t$.  We assume $Y_{it}$ follows a Binomial($T_i$, $\mu_{it}^O$) distribution. The period $t$ may be a week, month, or other duration depending upon study design. %
We assume the mean model
\begin{equation}
q(\mu_{it}^O) = \psi_s + \xi_i + \mathbf{z}_i^\mT\boldsymbol{\gamma} + {\bm h}_s(t,\nu_s)^\mT\boldsymbol\delta_s +  {\bm g}\left(x_{it}\right)^\mT\boldsymbol\beta_s,
\label{eq:mu-model}
\end{equation}
where $q(\cdot)$ is the logit link function.
 The $\psi_s$ are study-level  effects distributed as $N(0, \sigma^2_\Psi)$ 
 and the $\xi_i$ are subject-level random effects distributed as $N(0, \sigma^2_\xi)$. 
The vector $\mathbf{z}_i$ is a set of covariates with coefficients $\boldsymbol{\gamma} \sim N(0, \sigma^2_\Gamma)$.  The function ${\bm h}_s(t, \nu)$ accounts for temporal variation in disease risk via a flexible spline with $\nu$ degrees of freedom. As in  the exposure concentration model, we use a B-spline basis for time that is generated separately for each study and included in a block-diagonal fashion in the matrix formulation of the model. The time spline coefficients $\delta$ are modeled as $N(0, \sigma^2_\Delta)$.
The assigned PM exposure is denoted $x_{it}$, and $ {\bm g}\left(x_{it}\right)^\mT$ represents a transformation of these values using splines with coefficients $\boldsymbol\beta$. We discuss these in the following subsections.

\subsection{Assigning Individual Exposure: $x_{it}$}
\label{sec:exp_eta}
An exposure summary value derived from the posterior distributions of model \eqref{eq:full_exp_model} is needed for the outcome model as $x_{it}$.
Here we  set $x_{it}$ to be the long-term average indoor concentration for the household of individual $i$ over the last $\tilde{T}$ days.
We compute $x_{it}$ from the exposure model outputs as:
\[x_{it} = \frac{1}{\tilde{T}}\sum_{t' =t - \tilde{T} + 1 }^{t } \sum_g\sum_k I_{gkit'}\E[\eta_{gki}| \bm w],\]
where $I_{gkit'}$ is an indicator of subject $i$ being in stove group $g$ and cluster $k$ at time $t'$.
This averages the mean exposure concentrations from the $\tilde{T}$ times prior to $t$, excluding the temporal trend. This captures changes in stove group and cluster in cross-over studies, otherwise this value will be the same for an individual at all times.  The `washout' period implied by $\tilde{T}$ can vary, here we primarily consider a period of 28 days. 

The choice of a long-term average that excludes the time trend, as opposed to short-term exposure or other measures of long-term exposure, represents a compromise of several competing pragmatic and statistical objectives. 
There are a limited number (usually 1 to 3) of measurements per child and the temporal precision of the measurements is sometimes limited due to confidentiality concerns. This can make modeling at the daily or weekly scale difficult without strong assumptions about the temporal structure. We anticipate that the results will be used to inform future estimates of disease burden, which are typically computed on an annual-average time scale \citep{Burnett2014}. Smoothing over temporal differences to create a long-term average introduces Berkson error, but can reduce classical error compared to using the small number of measured values directly \citep{Carroll2006}. In the simulations below, we show how excluding the temporal trend from $x_{it}$ can reduce error in the estimated long-term average.

\subsection{Exposure-response curve}
\begin{figure}
\begin{center}
\includegraphics[width=0.7\textwidth]{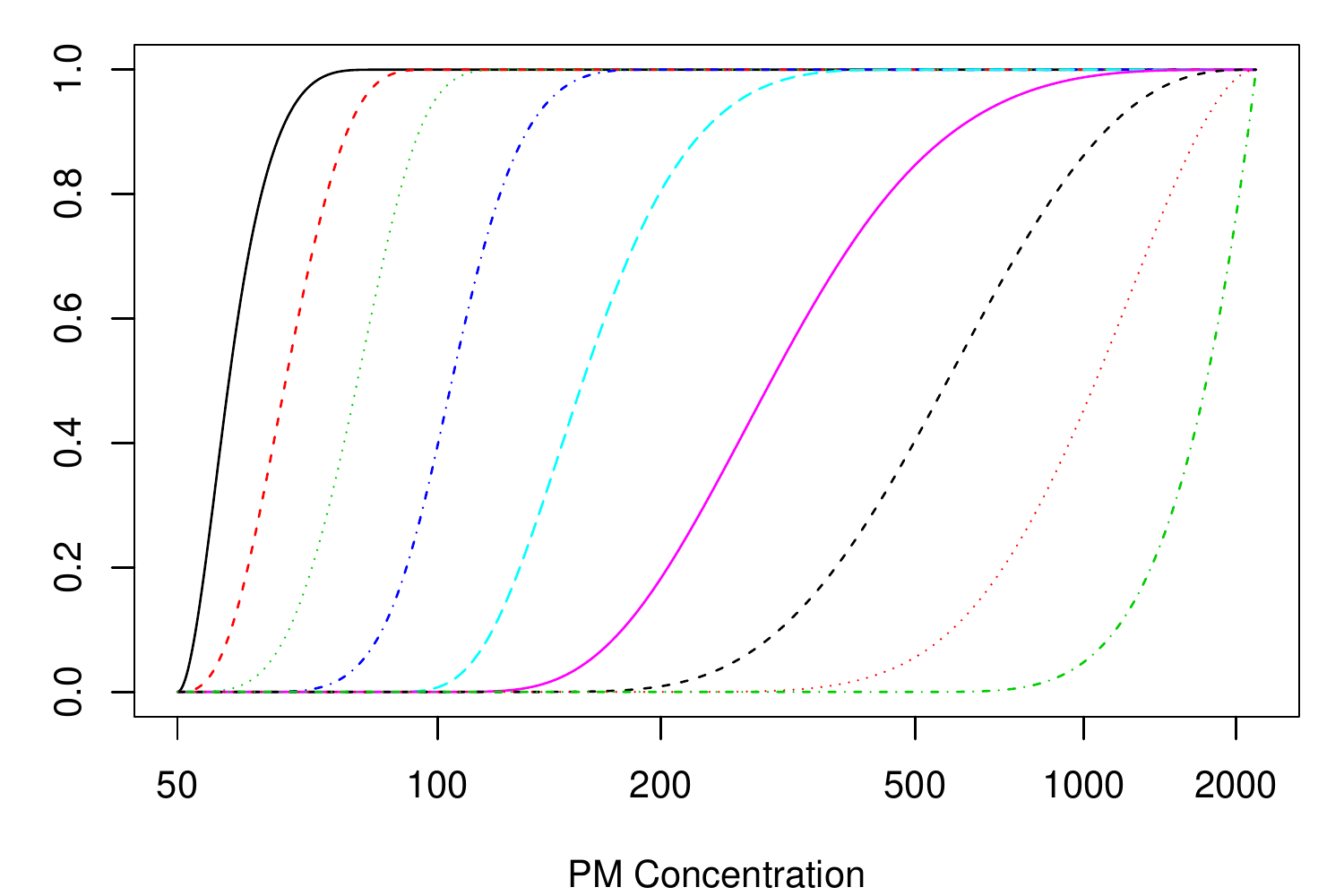}
\caption{I-Splines for the combined analysis of the three Nepal studies, using boundary knots at 50 and 2,200 $\upmu$g/m$^3$.} \label{fig:app:combined_model_splines}
\end{center}
\end{figure}

The function ${\bm g}(\cdot)$ represents a transformation of the assigned long-term exposure $x_{it}$ using splines and coefficients $\boldsymbol\beta_s$. 
I-splines provide a monotonic basis, which allows for estimating a non-decreasing exposure-response function when the coefficients are restricted to be non-negative \citep{Ramsay1988}.
This approach has been used before in an outdoor air pollution context by \cite{Powell2012}.
The splines require some choice of endpoints and interior knots, which can be tailored to %
each application. Figure~\ref{fig:app:combined_model_splines} shows the I-Spline basis functions used in the application in Section~\ref{sec:application_to_nepal_combined_outcome}.

For our  primary results, we use a single vector of coefficients $\boldsymbol\beta$  for all studies  to jointly estimate an exposure-response curve across the full range of data.
We also consider allowing $\boldsymbol\beta$ to differ by study by allowing
each study to have its own coefficient vector $\boldsymbol\beta_s$. The vector of the $j$th element of each study's coefficient vector is modeled: $(\beta_{1j}, \dots, \beta_{Sj})^\mT \sim N(\beta_0, \sigma_B^2\mathbf{V})$. 
The variance matrix $\mathbf{V}$  is modeled by decomposing it into a variance vector $\boldsymbol\xi_{0}$ and correlation matrix $\boldsymbol\Sigma$, i.e. $\mathbf{V} = diag(\boldsymbol\xi_{0})\boldsymbol\Sigma diag(\boldsymbol\xi_0).$ 
We consider $\boldsymbol\xi_0$ to be fixed, although a non-negative prior is possible. An `LKJ' prior is put on $\Sigma$, which provides  a way to parameterize the correlation matrix as deviations from the identity matrix indexed by a single hyperparameter \citep{Lewandowski2009}.

We plot estimated exposure-response curves using the posterior means of $\boldsymbol\beta$. We show the estimated pointwise uncertainty around the curve, including uncertainty about the study-level intercepts.

\subsection{Implementation}
We implement the exposure concentration model and the outcome model in the R package \texttt{bercs}, which stands for  ``Bayesian Exposure-Response Curves via STAN''. This R package is publicly available at  www.github.com/jpkeller/bercs.
  Sampling is done via the STAN modeling framework \citep{Stan2018}, which can efficiently sample from the posterior distributions of the parameters. The STAN framework allows for easy incorporation of the restrictions on model parameters so that, for example, the spline coefficients can be forced to be non-negative.  The No-U-Turn-Sampler used by STAN requires many fewer iterations (on the order of a few thousand, even for a complex model) than a traditional markov-chain monte carlo approach to generating a posterior sample.

\subsection{Accuracy of Approximation to Joint Exposure-Outcome Model} 
A complete Bayesian paradigm would jointly fit the exposure and outcome models, propagating uncertainty from all stages into the uncertainty of the exposure-response curve parameters. However, as noted above in the rationale for fitting separate exposure models by study, we do not wish to have exposure estimates from different studies influencing one another when they may not be comparable in accuracy, meaningfulness of clusters, and other factors. Additionally, fitting a joint model for exposure and outcome would greatly increase the number of parameters, leading to computational challenges in exploring the parameter space.

  The use of an exposure $x_{it}$ derived from the posterior mean $\E[\eta_{gki} | \bm w]$ is an approximation to passing the full posterior distribution of $\eta_{gki}$ into the health model. As a sensitivity analysis in the application, we explore the impact of this approximation by fitting models to a small sample of points from the exposure concentration posterior space and using them to estimate an exposure-response curve from the outcome model.

\section{Simulations}
\label{sec:simulation}
We demonstrate the flexibility and robustness of the model through a series of simulations. We first simulated data for the exposure model, and then used the data and estimates from the fitted exposure model to generate outcome data and fit the outcome model.

\subsection{Exposure Model Simulation}
We conducted a simulation of the exposure model to assess benefits from model shrinkage, in terms of reduced error in the exposure estimates, and sensitivity to the prior distributions.
We simulated exposure data based upon three different parallel designs, two with 2 groups and one with 4 groups, that were structured to be similar to the data from the motivating studies. Data were generated from the model $ w_{gkit} \sim N\left(\eta_{g} + \alpha_{0k} + \alpha_{1i}  + \tilde{f}(t), \sigma^2_{W}\right)$, where $\alpha_{1i} \sim N\left(0, \sigma_{H}^2\right)$ and $\alpha_{0k} \sim N\left(0, \sigma_{K}^2\right).$
This model uses fixed values of the group means $\eta_g$ and time trend $\tilde{f}(t)$ to generate the data, while the exposure model \eqref{eq:full_exp_model} that we fit uses splines to estimate the time trend and models the $\eta_g$ as drawn from a prior distribution. In simulation Setup 1, there were two groups with means 4 and 5, each comprised 12 clusters ($k=1, \dots, 12*2$) and within each cluster there were 50 households ($i=1, \dots, 2*12*50$), each with 2 observations.  Simulation Setup 2 had the same structure, except group means were 3 and 6, the amplitude of temporal variation was smaller, and the observation  variance was smaller. This structure was designed to be more favorable for using the observed values.
In simulation Setup 3, there were four groups with no clustering and each containing 200 households  with a single exposure measurement.
Section 2 of the Supplementary Material provides further details on the simulation settings.

We computed the error of estimates of the group means ($\mu_g = \eta_g$),  the cluster means ($\mu_{gk} = \eta_g + \alpha_{0k}$), and the household means ($\mu_{gki} = \eta_g + \alpha_{0k} + \alpha_{1i}$). Letting $\hat\mu^{(b)}$ denote an estimate of a particular mean  for simulation replication $b$, error for group means was assessed using mean squared error, $MSE = \frac{1}{B}\sum_{b=1}^B\left[\hat\mu^{(b)} - \eta_g\right]^2$, while error for the cluster and household means was assessed using mean squared prediction error $MSEP = \frac{1}{B}\sum_{b=1}^B \left[\hat\mu^{(b)} - \mu^{(b)}\right]^2$, where $\mu^{(b)}$ is the realization of $\mu_{gk}$ or $\mu_{gki}$ in simulation $b$. %
For each model level (group, cluster, household), we compute the MSE or MSEP using the observation averages ($\hat\mu_1$), the posterior mean exposure value including time trend ($\hat\mu_2$), and the posterior mean exposure value excluding time trend ($\hat\mu_3$). The formulas for each estimator for each mean is provided in  Section 2.2 of the Supplementary Material.

Table~\ref{tab:expsim_etag_metrics} shows the mean squared errors for the different estimates of group, cluster, and household means. 
For estimating the group means, the approaches that included time variation ($\hat\mu_1$ and $\hat\mu_2$) had  less error than the posterior mean that excluded time ($\hat\mu^3$), although in Setup 2 this difference is small.  The opposite trend occurs when evaluating household means, which is the exposure measure of interest. In that case, the modeled values  do better than the observed values, and the lowest error occurs when excluding the time trend.
This demonstrates the importance of using shrinkage to improve estimation when there are limited number of observations for each household.

\begin{table}
\centering
\caption{MSE and MSEP for estimates of long-term exposure at different model levels from the simulations.}
\label{tab:expsim_etag_metrics}
\begin{tabular}{ccccc}
  \hline
  & & Group & Cluster & Household \\
Setup & Estimator &   Means ($\mu_g$) &  Means ($\mu_{gk}$) &  Means ($\mu_{gki}$)\\
\hline
1& $\hat{\mu}^1$		 & 0.001 & 0.016 &  0.75 \\
			& $\hat\mu^2$ & 0.001 & 0.015 & 0.34 \\
			& $\hat\mu^3$				 & 0.015 & 0.017 & 0.09 \\
2 &  $\hat{\mu}^1$	 & 0.009 & 0.009 & 0.375 \\
 &  $\hat{\mu}^2$	& 0.009 & 0.009 & 0.304 \\
 &  $\hat{\mu}^3$	 & 0.011 & 0.006 & 0.056 \\			
3 & $\hat{\mu}^1$			 	& 0.007 & -- &  1.50 \\
			& $\hat{\mu}^2$ & 0.007 & -- & 0.72 \\
			& $\hat{\mu}^3$			 & 0.024 & -- & 0.24 \\
   \hline
   
\end{tabular}
\end{table}

We also conducted simulations that reduced the number of observations per household to 1 under Setup 1 and that use priors more closely concentrated around 0. Results from those sensitivity simulations are provided in {Table 2} of the Supplementary Materials. Similar to the primary results, the shrinkage estimates performed the best for estimating long-term household averages.

\subsection{Outcome Model Simulation}

\begin{figure}[tb]
\begin{center} 
\includegraphics[width=0.95\textwidth]{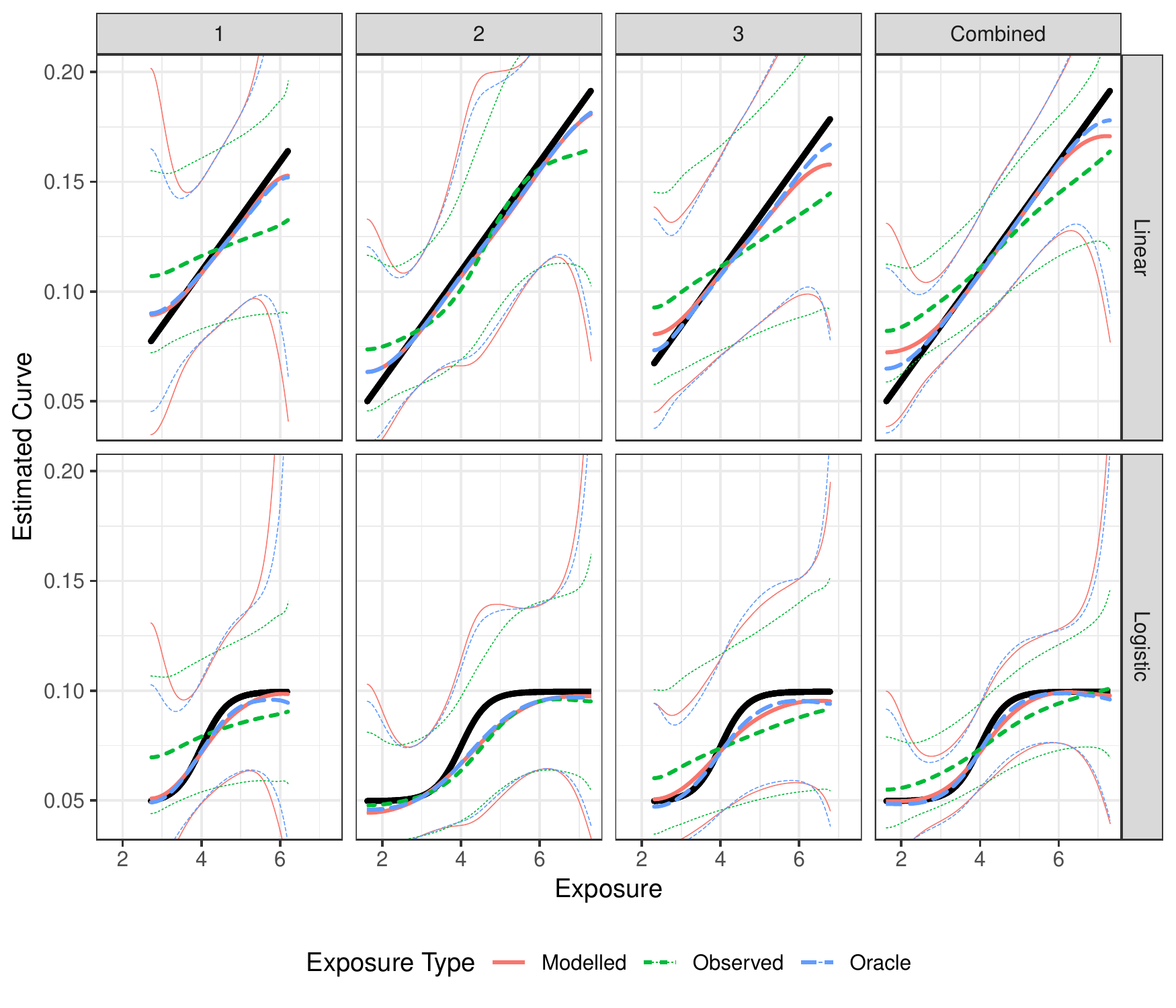}
\caption{Data generating (heavy solid line) and fitted curves for the outcome model simulation. Panel columns represent different setups, panel rows correspond to different exposure-response functions. Pointwise 95\% credible intervals are shown as lightweight lines.}
\label{fig:outsim_erc_fits}
\end{center}
\end{figure}

For the outcome model simulation, we used  exposure values from the exposure simulations. We generated data assuming constant time at risk with $q(\mu_{it}^O) = \psi_S + \xi_i + \tilde{h}(t) +  \tilde{g}(x_{it})$,
where we fixed the time trend $\tilde{h}(\cdot)$ to be a cosine function with amplitude and period 1. On the odds scale, we considered a linear form ($\tilde{g}(x) = \log [1 + 0.5(x- \log(5))]$)
 and a logistic form ($\tilde{g}(x) = \log[1 + 1/(1+\exp[-3(x-4)])]$) for the exposure response curve.
These functions are shown as the heavy solid curves in Figure~\ref{fig:outsim_erc_fits}.

The binary outcomes were generated with $\psi_S = -3$, corresponding to background odds of 0.05, and using the true long-term averages from the exposure simulation ($\eta_{gki}$). The primary model  used the estimated long-term exposures from the exposure simulation ($\E[\eta_{gki}|\bm w]$) for model fitting. We also compared against using the true exposure values ($\eta_{gki}$) as an oracle and using the average of the exposure measurements ($\sum_t w_{gkit} $), which we expect to have a large amount of measurement error due to both the unaccounted  temporal variation in the timing of the observations and the lack of shrinkage. We fit four sets of outcome simulations using the observations from Setup 1, Setup 2, Setup 3, and a model that combines  all three.

For each simulation, we estimated the exposure response curve %
using a common set of knots to facilitate averaging results using basis functions that are the same across replications.  We fit models with no restrictions on the $\beta$ and models that force the $\beta$'s to be non-negative.
Similar to \cite{Powell2012},  we evaluated the estimated curves pointwise at different exposure levels ($x_\ell$) using relative mean bias 
and root mean squared error (RMSE) evaluated across each replication $b$: 
\begin{align*}
\text{Relative Bias}(x_\ell) &= \frac{1}{B}\sum_{b=1}^B \left[\left(\hat\psi_S + {\bm g}\left(x_{it}\right)^\mT\boldsymbol\beta_s^{(b)} \right) - \left(\psi_S + \tilde{g}(x_{\ell})\right)\right] \Big / \left[\psi_S + \tilde{g}(x_{\ell})\right]\\
RMSE(x_\ell) &= \sqrt{\left(\frac{1}{B}\sum_{b=1}^B \left(\left(\hat\psi_S + {\bm g}\left(x_{it}\right)^\mT\boldsymbol\beta_s^{(b)} \right) - \left(\psi_S + \tilde{g}(x_{\ell})\right)\right)^2\right)}
\end{align*}
We include the intercept in these measures of performance so that the estimated curves are centered.

\begin{figure}[tb]
\begin{center} 
\includegraphics[width=0.95\textwidth]{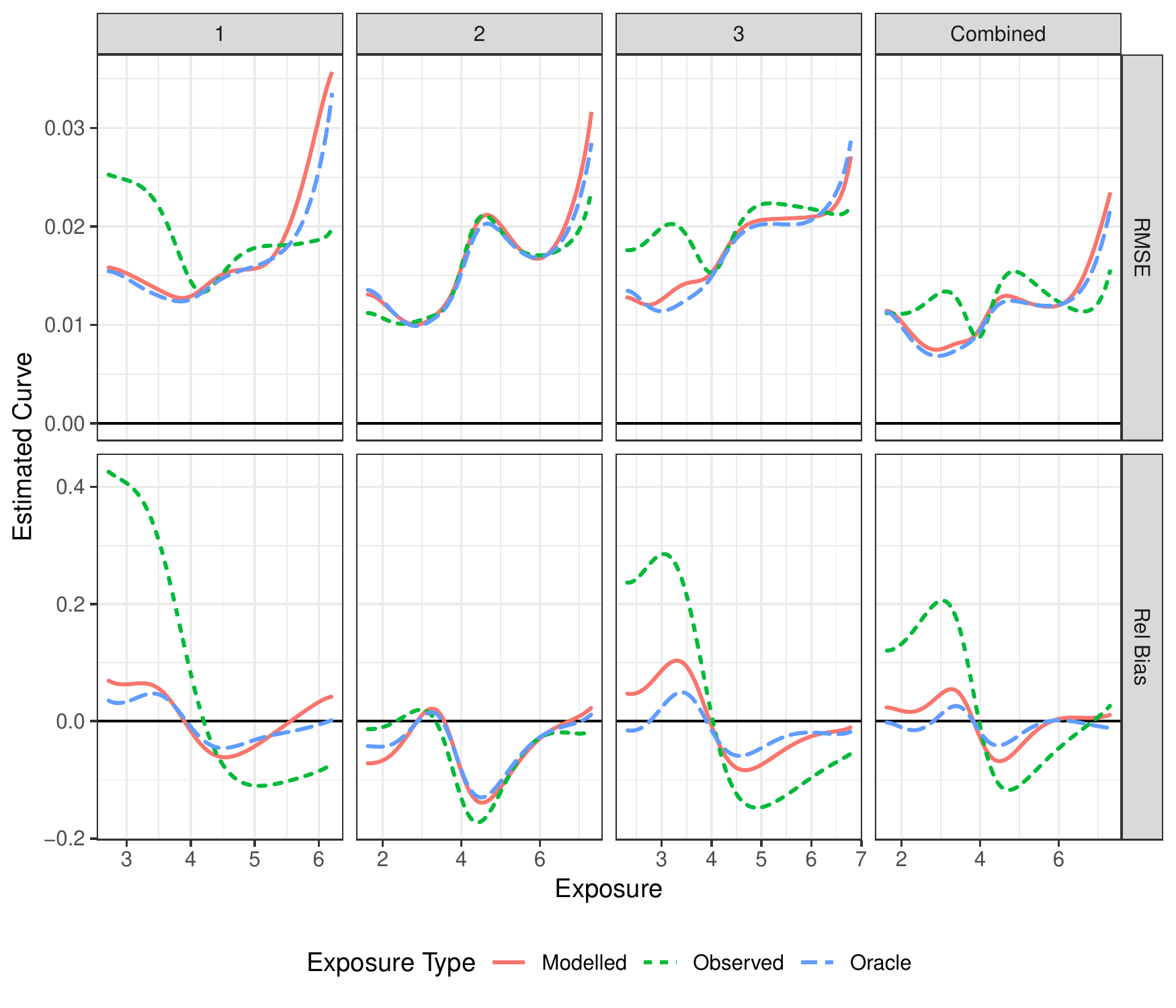}
\caption{Measures of exposure curve error (top row) and relative bias (bottom row) for the logistic exposure-response curve.}
\label{fig:outsim_erc_metrics_E4}
\end{center}
\end{figure}

The estimated curves are plotted in Figure~\ref{fig:outsim_erc_fits}. The mean estimated exposure-response curves derived from the modeled exposures come closer to matching the data-generating exposure-response function than the curves derived from the observed exposures in all settings, although the difference is only slight in Setup 2.  
The curves based on modeled exposures perform only slightly worse than the curves based upon the true (oracle) exposure values (Figure~\ref{fig:outsim_erc_metrics_E4}). Despite the notable bias, the curve based on observed exposures is less variable overall, leading to smaller RMSE than the other curves at highest exposure values (Figure~\ref{fig:outsim_erc_metrics_E4}).
The benefit of combining studies can be seen in the reduction in RMSE in the combined results  compared to the RMSE of the models from individual setups.  Additional benefit is seen in the reduced uncertainty for the exposure-response curve fit to the combined data (Figure~\ref{fig:outsim_erc_fits}).

\section{Application to Nepal Studies}
\label{sec:application_to_nepal}
We now present results from applying these models to the data from the motivating studies. We first present results for the exposure model applied to each study. %
The primary results for a combined analysis of the exposure-response curve are in Section~\ref{sec:application_to_nepal_combined_outcome}. Details for the chosen prior distributions are provided in Section 3 of the Supplementary Material.

\subsection{Exposure Models}

In the Bhaktapur Study, no
 information was available about household clustering, so no cluster-random effect was included. 
The posterior means of the stove group average concentrations (Table~\ref{tab:exp_stove_postmeans}) show that the households using an electric stove had the lowest PM concentrations, while those using biomass stoves the highest. 
The household random effect standard deviation ($\sigma_H$) had a small posterior mean (0.08), due to the influence of the prior and each household having only one observation (Table~\ref{tab:exp_variance_means}). This lead to a large amount of shrinkage in the fitted values, which are plotted in Figure~\ref{fig:expdata}. The pooling factors presented in Table~\ref{tab:exp_variance_means} show this as well; the pooling factor of 0.98 for the household level reflects the near-complete pooling at that level.

\begin{table}[t]
\begin{center}
\caption{Posterior means and 95\% credible intervals  for the long-term average PM concentration  for stove groups means in each study.  Estimates are provided on 
the original scale ($\exp(\eta_g)$; units of $\upmu$g/m$^3$).}
\label{tab:exp_stove_postmeans}
\begin{tabular}{l c}
 Stove Group  & $\exp(\eta_g)$\\
\hline
\textbf{Bhaktapur}\\
 Biomass & 387 (347, 437) \\
Kerosene & 119 (106, 133) \\
Gas 	& 72 (65, 81) \\
Electricity & 56 (49, 62)\\
\textbf{Sarlahi Phase I }\\
 Traditional Biomass & 1118 (934, 1312)\\
		 Improved Biomass & 584 (507, 672) \\
\textbf{Sarlahi Phase II}\\
 Improved Biomass & 395 (311, 503) \\
		 LPG & 	186 (148, 237) \\	
\hline
\end{tabular}
\end{center}
\end{table}

\begin{table}[t]
\begin{center}
\caption{Posterior means  (95\% credible intervals) for standard deviations and estimated pooling factors are each level of the exposure model.}
\label{tab:exp_variance_means}
\begin{tabular}{l  c c c}
& Posterior Mean & Pooling\\
Parameter & (95\% CI) & Factor\\ 
\hline
Bhaktapur \\
Household: $\sigma_H$ & 0.08 (0.00, 0.21) & 0.98 \\
Observation: $\sigma_W$  & 0.81 (0.77, 0.85) & 0.02 \\
Sarlahi Phase I  \\
Cluster: $\sigma_K$ & 0.21 (0.16, 0.26) & 0.19 \\
Household: $\sigma_H$ &  0.28 (0.24, 0.32) & 0.79 \\
Observation: $\sigma_W$ & 0.72 (0.70, 0.73) & 0.12  \\
Sarlahi Phase II \\
Cluster: $\sigma_K$ & 0.20 (0.12, 0.29) & 0.54 \\
Household: $\sigma_H$ & 0.08 (0.00, 0.22) & 0.99 \\
Observation: $\sigma_W$ & 0.91 (0.87, 0.95) & 0.03\\
\hline
\end{tabular}
\end{center}
\end{table}

 The modeled means for Phase 1 of the Sarlahi study are shown in Figure~\ref{fig:expdata}. The exponentiated posterior means of the stove-group intercepts were 1,118 $\upmu$g/m$^3$ and 584 $\upmu$g/m$^3$ for the traditional and improved biomass stove, which represents an approximate 50\% reduction in PM${2.5}$
 (Table~\ref{tab:exp_stove_postmeans}). The posterior means of the standard deviations of the cluster-level  and household-level random effects were similar in magnitude, 0.21 and 0.28 respectively (Table~\ref{tab:exp_variance_means}). The pooling factor of $\lambda = 0.19$ for the cluster level indicates a small amount of  shrinkage towards the random effect mean of zero. At the household level, the larger pooling factor of $0.79$ indicates a large amount of pooling towards the cluster-level random effects, but still some individual variation compared to the Bhaktapur study, which had very small random effects and a pooling factor of 0.98.

For Phase 2, the posterior mean PM$_{2.5}$ concentrations, after transformation back to the original scale, were 395 $\upmu$g/m$^3$ and 186 $\upmu$g/m$^3$ for the improved biomass and LPG stove groups, respectively (Table~\ref{tab:exp_stove_postmeans}). The mean LPG concentration was higher than that in the Bhaktapur study, which could be due to many factors including the use of other biomass stoves within the household. The standard deviation of the cluster random effect had a posterior mean of 0.20, more than twice the value of the posterior mean for the household random effect standard deviation (0.08) (Table~\ref{tab:exp_variance_means}). This leads to almost complete pooling at the household level ($\lambda = 0.99$) and moderate amount of pooling at the cluster level ($\lambda = 0.54)$. The deviations of some clusters from the group means can be seen in Figure~\ref{fig:expdata}, which also shows the large amount of shrinkage overall.

\subsection{Outcome Models -- Study Specific Analyses}
\label{sec:application_to_nepal_single_outcome}
\begin{figure}[t]
\begin{center}
\includegraphics[width=\textwidth]{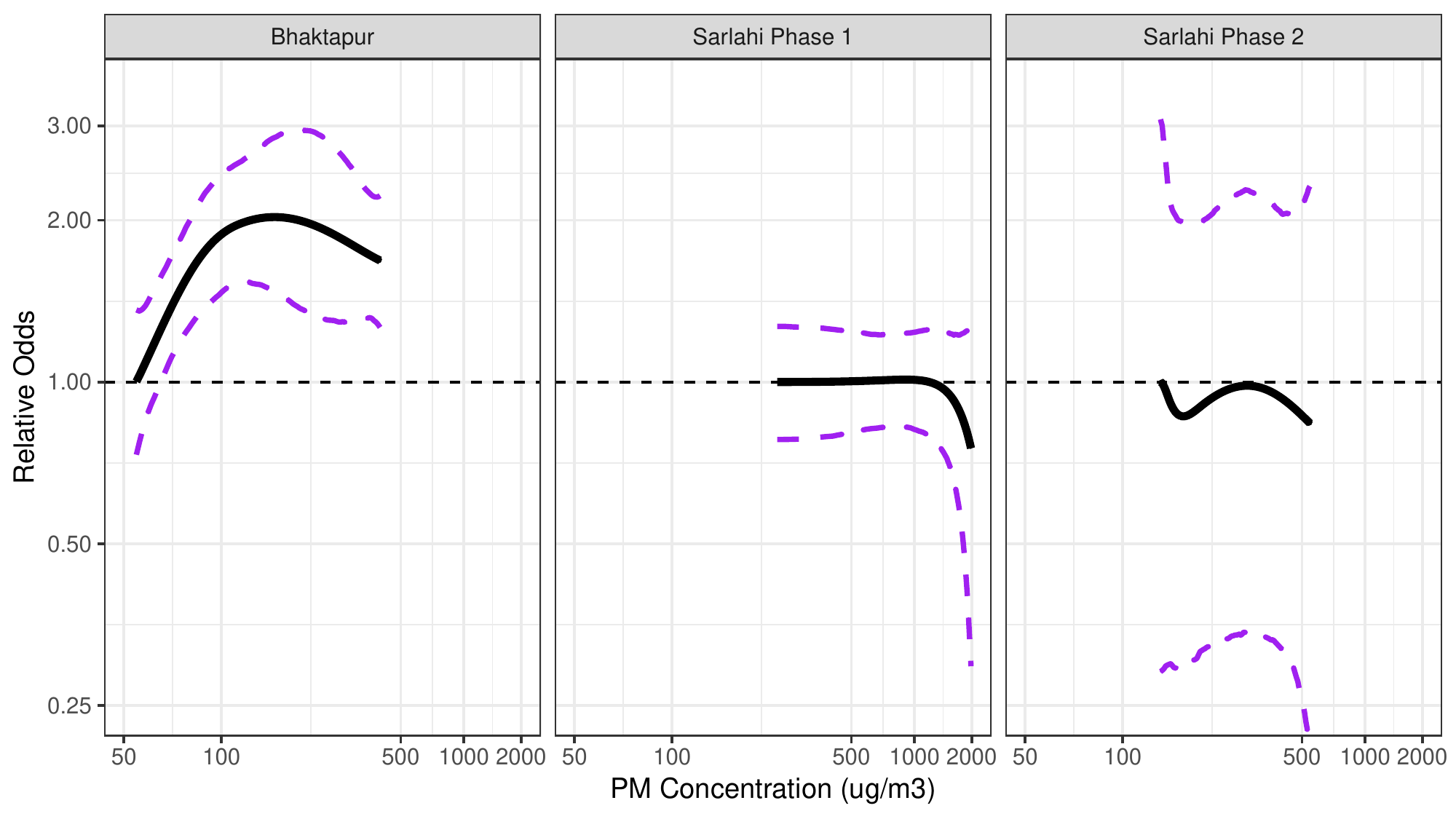}
\caption{Estimated exposure-response curve for the three studies.}
\label{fig:out_fit_individ}
\end{center}
\end{figure}

 Because of the case-control design of the Bhaktapur study, we cannot directly estimate the risk rate for ALRI, however we are able to estimate the relative odds of ALRI across different long-term PM$_{2.5}$ concentrations (Figure~\ref{fig:morb_comb}).  
The fitted exposure-response curve is plotted in the left panel of Figure~\ref{fig:out_fit_individ} and shows a large increase in odds from approximately 50 $\upmu$g/m$^3$ to 100$\upmu$g/m$^3$. This represents an almost doubling of ALRI odds for subjects in the gas, kerosene and biomass stove groups relative to those in the electricity group.  This is consistent with the results obtained by \cite{Bates2018}, who also adjusted for additional covariates in their outcome model.

For computing long-term exposures $x_{it}$ in Phase 1 of the Sarlahi study, we set $\tilde{T} = 28$ days. This allows for a piece-wise constant transition in the estimated long-term average exposure concentrations from the traditional stove to the biomass stove. This time period is also short enough that it provides for meaningful contrasts in exposure in the amount of follow-up time available after stove intervention, which ranged from 6 to 12 months. A $N_+(0, 1)$ prior distributions was used for  $\sigma_\gamma, \sigma_\beta, \sigma_I, \sigma_K$, and $\sigma_\delta$. A natural cubic spline with 8 $df$ was used to account for temporal variations in ALRI risk. Because of the step-wedge design, this limits the information available  for estimating the exposure-response effect to the period in which only part of the cohort had received the improved stoves.

From a model that adjusted for time but did not include PM concentration, the posterior mean associated with an indicator of traditional biomass stove was 0.08 ($-0.04$, 0.22). This corresponds to a reduction of 7.7\% ($-4.0$\%, 19.7\%) in odds of ALRI in the improved biomass stove group compared to the traditional stove group. When including PM concentration (but not stove type) and fitting the full outcome model, there remained little evidence of an exposure-response relationship (Figure~\ref{fig:out_fit_individ}, middle panel). This was robust to different choices in the number and location of knots (data not shown).

For Phase 2,  the relative difference in the unadjusted rates show a reduction in the LPG group (4.58 versus 3.52 cases per 100 person-years), but the limited amount of information yields no evidence of  a dose-response relationship (Figure~\ref{fig:out_fit_individ}, right panel).

\subsection{Outcome Model -- Combined Analysis}
\label{sec:application_to_nepal_combined_outcome}
For our primary model that combines all three studies together, we set the I-spline boundary knots at (the logarithm of) $50$ and $2,200$ $\upmu$g/m$^3$  and the interior knots at (the logarithm of)  60, 85, 100, 125,  200, and 500 $\upmu$g/m$^3$ (Figure~\ref{fig:app:combined_model_splines}). This allows for most of the spline flexibility to be in the lower concentrations, where the most overlap across stove groups occurs.

\begin{figure}[t]
\begin{center}
\subfloat[][]{\includegraphics[width=0.48\textwidth]{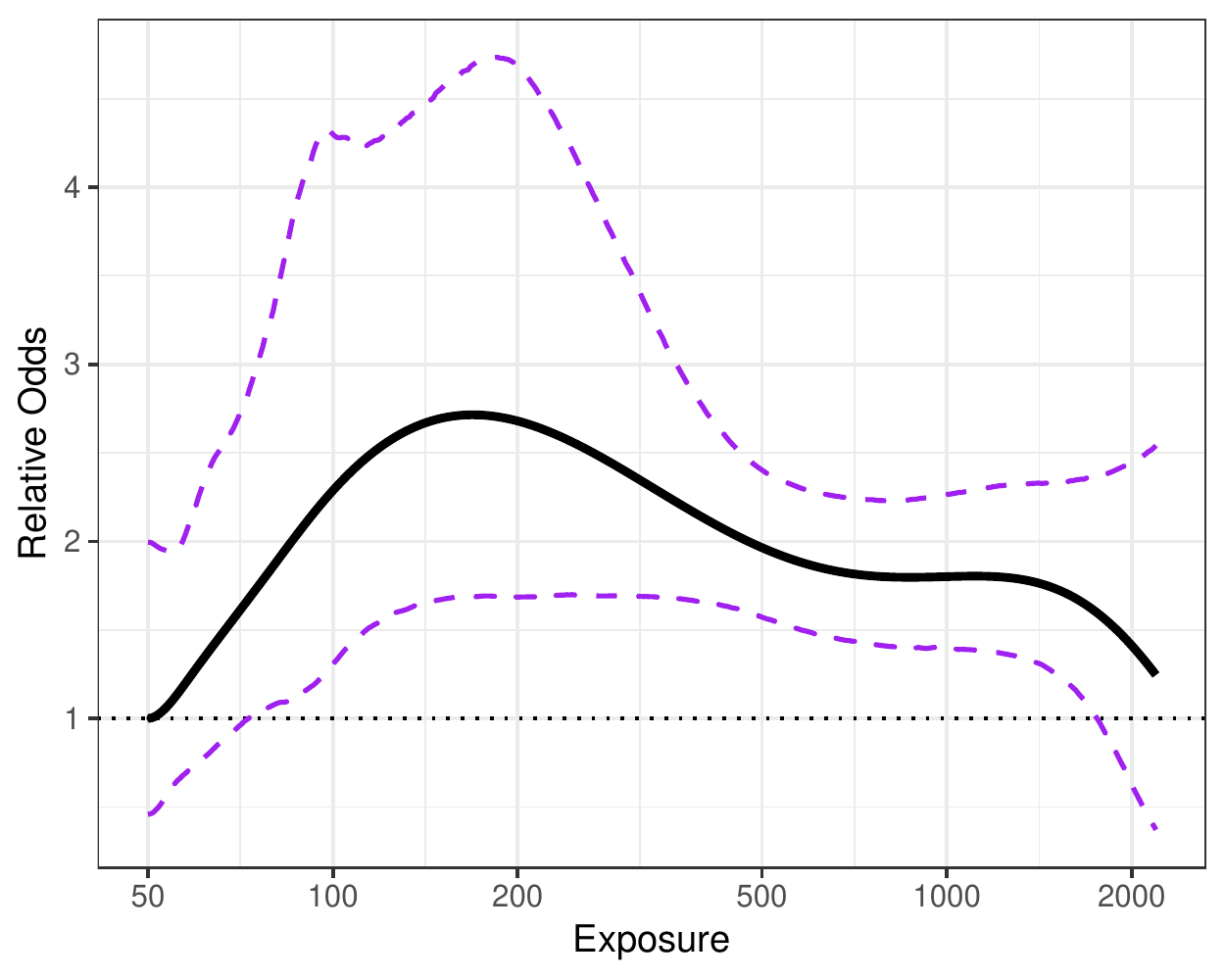}\label{fig:combined_fit}}
\subfloat[][]{\includegraphics[width=0.48\textwidth]{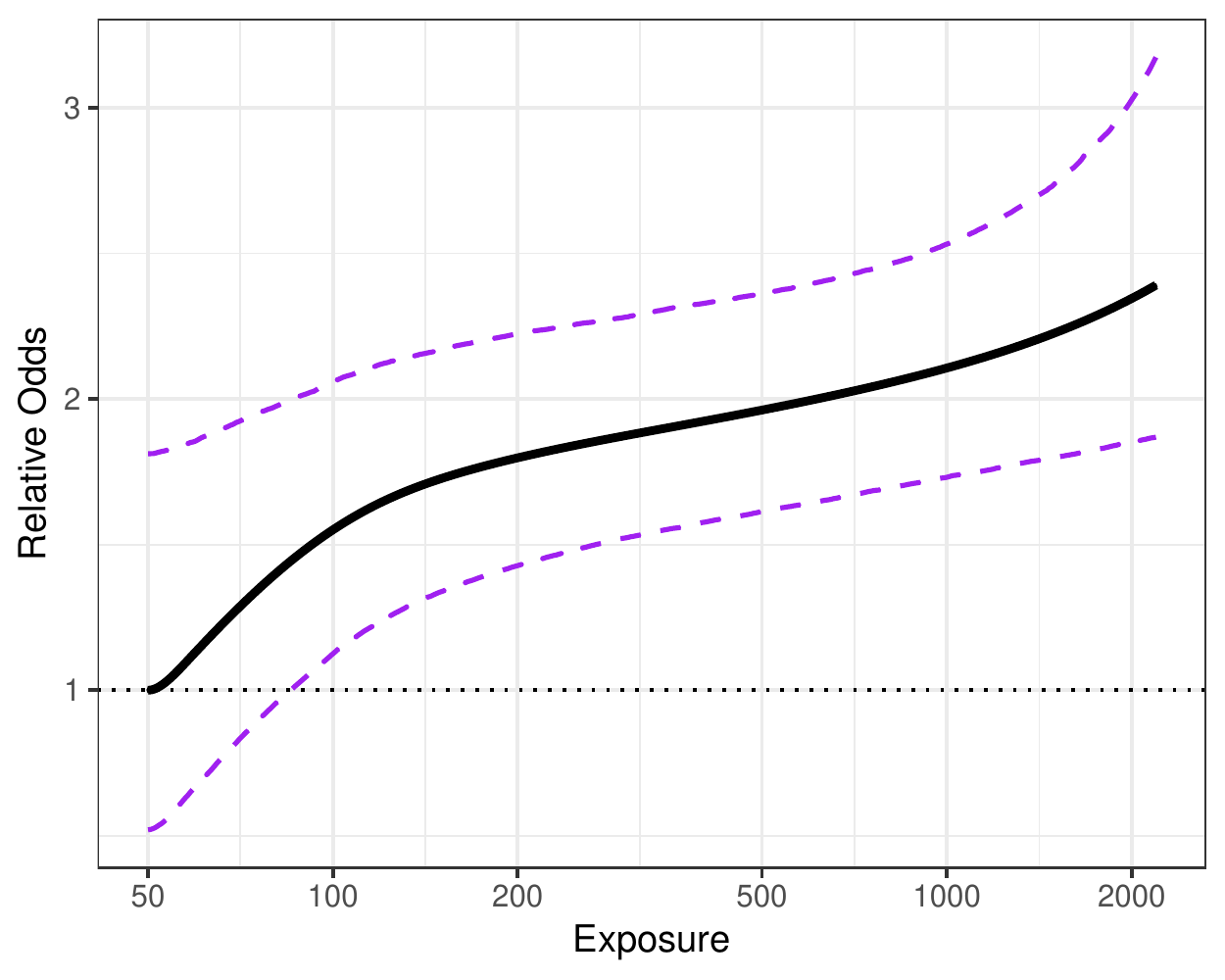}
\label{fig:combined_restrictBeta_fit}}
\caption{Estimated exposure-response curve for all three studies combined. Panel (a) places no restriction on spline coefficients, (b) restricts the curve to be non-decreasing.}
\end{center}
\end{figure}

Figure~\ref{fig:combined_fit} shows the estimated exposure-responsive curve from the joint model. The overall shape is similar to the Bhaktapur study curve (Figure~\ref{fig:out_fit_individ}, left panel), but the inclusion of the Sarlahi studies extends the horizontal axis out to 2200 $\upmu$g/m$^3$. This curve is consistent with the lack of evidence for an exposure-response relationship in the Sarlahi study models, since those studies only included PM concentrations of approximately 150 $\upmu$g/m$^3$ and higher. We see from the combined-model curve that the estimated dose-response relationship plateaus for concentrations above 500 $\upmu$g/m$^3$ and then drops at the highest concentrations where there is less data. Based upon the combined model fit, we would not expect to see an exposure-response relationship at the concentrations observed in the Sarlahi study. The estimated curve has a large amount of pointwise uncertainty for exposures between 100 and 300,  which is due in part to the contrasting evidence from the Bhaktapur study (which saw elevated odds for exposures in this range) and Phase 2 of the Sarlahi study (which did not see changes in odds of ALRI across this range).

When forcing the exposure-response curve to be non-decreasing (by requiring $\beta > 0$), we obtain the curve shown in Figure~\ref{fig:combined_restrictBeta_fit}. This follows the same basic trend as the unrestricted curve at low exposure levels, but continues to increase at higher exposure concentrations while the unrestricted curve decreases then plateaus.  However, the uncertainty around this curve is quite consistent with an increase in risk at low levels, followed by constant risk for concentrations above 200 $\upmu$g/m$^3$.

In the model that combines data from all three studies, we see a small shift in the study-specific intercepts (see Table~\ref{tab:outcome_variance_postmeans}). The standard deviation of the subject-level random effect is 1.32 (1.19, 1.46), close to the estimate from the Sarlahi Phase 1 model and notably higher than in the Bhaktapur or Sarlahi Phase 2 single-study models.

\begin{table}[t]
\begin{center}
\caption{Posterior means (95\% credible intervals) for outcome model parameters.}
\label{tab:outcome_variance_postmeans}
\begin{tabular}{l l c}
& & Posterior Mean \\
Study & Parameter & (95\% CI)\\
\hline
\textbf{Single Study}\\
Bhaktapur & $\psi_S$ & $-0.43$ ($-0.71$, $-0.16$)\\
		& $\sigma_I$ & 0.13 (0.01, 0.30) \\
Sarlahi Phase I & $\psi_S$ & $-8.76$ ($-9.01$, $-8.52$)\\ %
		& $\sigma_I$ & 1.32 (0.07, 1.46) \\
Sarlahi Phase II & $\psi_S$ & $-9.25$ ($-9.98$, $-8.71$) \\ %
		& $\sigma_I$ & 0.59 (0.03, 1.32) \\
\textbf{Pooled}\\ 
Bhaktapur & $\psi_S$ & $-0.70$ ($-1.39$, $-0.14$) \\
Sarlahi Phase I & $\psi_S$ & $-9.26$ ($-10.04$, $-8.55$)\\
Sarlahi Phase II & $\psi_S$ & $-10.90$ ($-11.84$, $-10.08$)\\
All Studies & $\sigma_I$ & 1.32 (1.19, 1.46) \\
\hline
\end{tabular}
\end{center}
\end{table}

As a sensitivity analysis, we fit an ALRI model that combined data from all studies, but allowed the estimated exposure-response curve to vary between studies. This pools information across studies for estimating the subject-level random effects and defining the exposure basis functions (which span the exposure levels seen in all studies). There is some pooling of information between studies for the exposure-response curve coefficients, due to the LKJ prior on the $\beta$'s.
Figure~1  in the Supplementary Materials shows the estimated curves from this model fit. The plotted curves do not include the study-specific intercepts, so they are artificially fixed at a relative odds of 1 on the left-hand side.  The estimated curves in  Figure~1 in the Supplementary Materials are all consistent with the combined-study curve in Figure~\ref{fig:combined_fit}.
But because the range of exposure values within each study only covers a portion of the overall exposure variability between studies, the uncertainty in the exposure-response curves is quite large when allowing the curve to vary by study.

\begin{figure}[t]
\begin{center}
\includegraphics[width=0.68\textwidth]{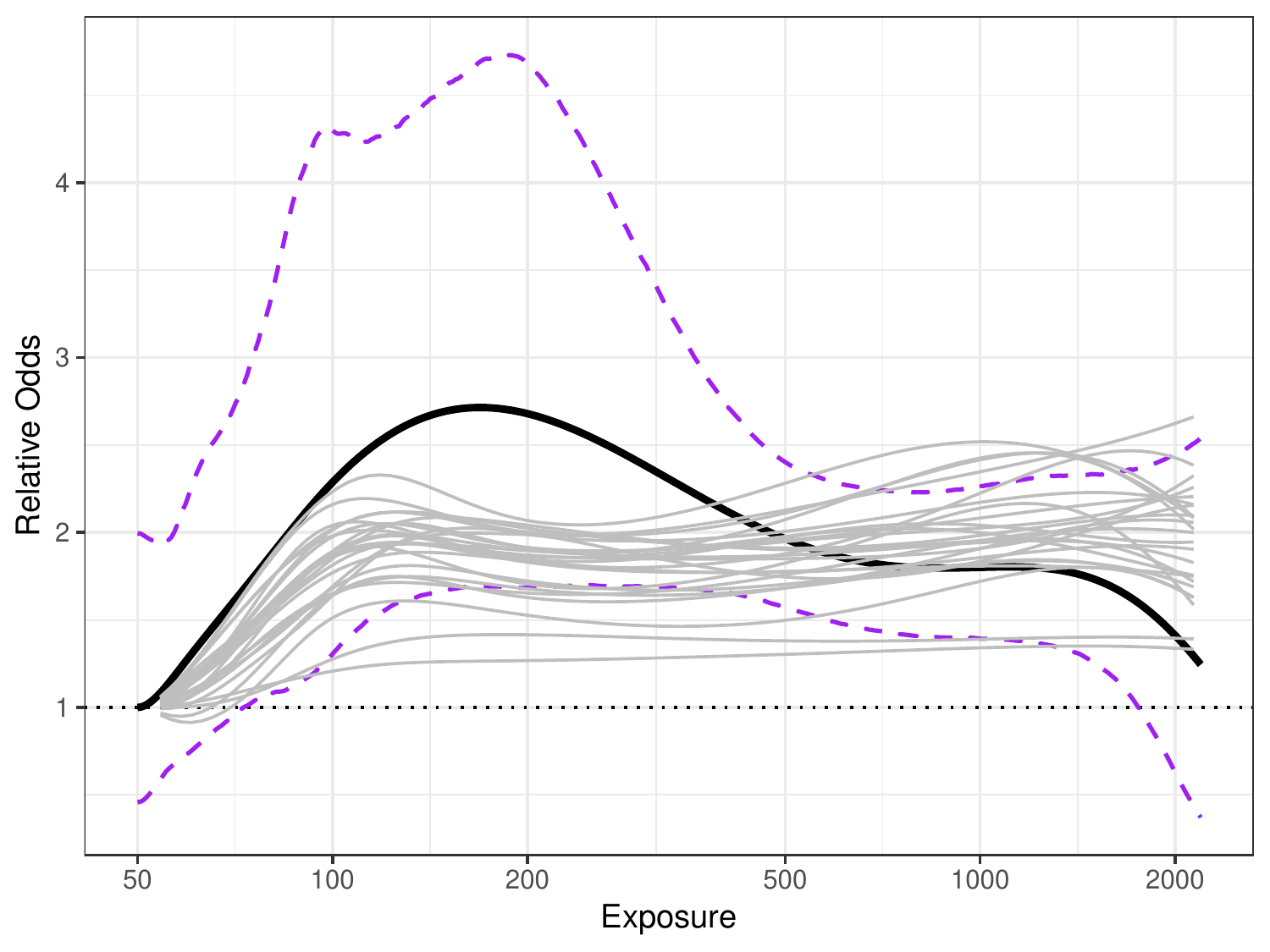}
\caption{Estimates exposure-response curves for 25 different draws of exposure model parameters from their joint posterior distribution. The blue curve is the main result, which uses the posterior mean of the exposure model parameters.}
\label{fig:comb_out_resampexp_fit}
\end{center}
\end{figure}

\subsection{Impact of exposure parameter variability on outcome}
As a sensitivity analysis, we fit the combined exposure-response curve model using exposure model parameters that were draws from their respective posterior distribution, instead of using the posterior mean.
Figure~\ref{fig:comb_out_resampexp_fit} shows the estimated curve for 25 different draws. This provides a representation of the uncertainty contained within the exposure model posterior. Relative to the primary result, which uses the posterior mean $\E[\eta_{gki} | \bm w]$, the sample draws yield curves that are slightly attenuated in the range of 100 to 500 $\upmu$g/m$^3$. However, the overall trend remains the same.

\section{Discussion}
We have presented exposure-concentration and exposure-outcome models that provide a flexible paradigm for combining highly-variable data from multiple studies in order to estimate an exposure-response curve. This model allows for irregularly-collected exposure measurements and provides a structure for pooling together estimates from different levels. The flexible, spline-based approach to estimating the exposure-response curve avoids restrictive assumptions required for a parametric curve and allows for varying uncertainty across the range of exposure values.  The framework can also easily accommodate different approaches to defining long-term exposures (e.g. including the time trend or modifying the averaging period) for different applications.

In the exposure models fit to the Sarlahi and Bhaktapur studies, we found a large amount of pooling towards group-level means. This reflects the large variance in household-level measurements, even within the same village or community. The pooling towards group-level mean can improve the power in the outcome models by eliminating excess noise in the exposure values. However, it also reduces the available exposure contrasts to approximately the number of stove groups. This is one important advantage for pooling studies: although the exposure values are primarily concentrated around a small number stove group means, the stove group means can still cover a large range of exposure concentrations across all studies.

The step-wedge design of Phase 1 of the Sarlahi Study presents several challenges for inference. First, the large time trends in ALRI rates (Figure~\ref{fig:morb_comb}) mean that only the cross-sectional contrasts between households provides information about exposure-response relationship. Furthermore, the lack of overlap between stove types in exposure measurements (other than a handful of locations which did not receive the improved biomass stove) means that the difference in exposure concentrations between stove types  is confounded by temporal trends in the exposure values. We mitigated this in the exposure model by using a single, smooth time trend so the apparent jump in the middle of the Sarlahi Phase I data in Figure~\ref{fig:expdata} represents the stove effect on concentrations.

The case-control design of the Bhaktapur Study prevented the direct assessment of ALRI risk in that cohort. However, the inclusion of controls allowed for the estimation of a relative-odds curve. This relies on the study-specific intercept terms in the outcome model we have presented, which estimates the overall odds of ALRI and allows only relative contrasts to inform the exposure-response curve. 

The modeling framework we have presented can flexibly include multiple studies and  future applications to cookstove intervention studies  can incorporate many more than just the three studies presented here. This will allow for continued refinement of the exposure-response curve using available evidence, which can then be used to target and plan future interventions to reduce ALRI risk.

\section*{Acknowledgements}
Funding for the Bhaktapur study was provided by the European Commission (EUeINCOeDC contract no. INCO-FP6-003740), the Danish Council of Developmental Research (project no. 91128), the Research Council of Norway (RCN project nos. 151054 and 172226), the Norwegian Council of Universities' Committee for Development Research, Education (NUFU project no. PRO 36/2002), and the Winrock International, Kathmandu, Nepal.
Funding for the Sarlahi studies was provided by the National Institute for Environmental Health Sciences (NIEHS) R01ES015558, The Thrasher Research Fund (9177), and the Global Alliance for Clean Cookstoves, UN Foundation (UNF-12-380). The Sarlahi studies were registered on Clinicaltrials.gov \# NCT00786877. 

Supplemental Material is available on request.

\bibliographystyle{apalike}
\bibliography{ERC_Manuscript_arXiv}

\end{document}